\documentclass[pra,twocolumn,showpacs]{revtex4}

\usepackage{epsfig}
\usepackage{graphicx}
\usepackage{bm}
\usepackage{amsmath}
\usepackage{dcolumn}

\newcommand{\rcm}{\mbox{cm$^{-1}$}}
\newcommand{\Sch}{Schr\"{o}dinger}

\begin{document}

\title{Experimental study of the Ca$_2$ $^1$S$+^1$S  asymptote}

\author{O. Allard}
\author{C. Samuelis}
\author{A. Pashov}\altaffiliation{On leave from the Institute for Scientific Research in Telecommunications,
 ul. Hajdushka poliana 8, 1612 Sofia, Bulgaria}
\author{H. Kn\"ockel}
\author{E. Tiemann}
\affiliation{Institut f\"ur Quantenoptk, Universit\"at Hannover,
  Welfengarten 1, 30167 Hannover, Germany}

\date{\today}

\begin{abstract}
The filtered laser excitation technique was applied for measuring
transition frequencies of the Ca$_2$ B-X system from asymptotic
levels of the X$^1\Sigma_{\mathrm g}^{+}$ ground state reaching
$v''=38$. That level has an outer classical turning point
of about 20~\AA\ which is only 0.2 \rcm\ below the molecular
$^1$S$+^1$S asymptote. Extensive analysis of the spectroscopic data,
involving Monte Carlo simulation, allowed for a purely
experimental determination of the long range parameters of the potential
energy curve. The possible values of the s-wave scattering length could be limited to be
between 250$a_0$ and 1000$a_0$.

\end{abstract}


  \maketitle

\section{Introduction}
 \label{intro}

In our previous paper \cite{Allard:02} we reported on an accurate
determination of the Ca$_2$ X$^1\Sigma_{\mathrm g}^{+}$ ground
state potential energy curve (PEC) from Laser Induced
Fluorescense (LIF) spectroscopy. The study was motivated by the
rapid progress in the development of the Ca frequency standard
\cite{Wilpers:02} and the recent photoassociation spectroscopy
data on the $^1$S$_0 + ^1$P$_1$ asymptote \cite{Zinner:00}. It was
considered as the first step of a comprehensive investigation,
which should provide a description of the X state for short and
intermediate internuclear distances, thus enabling the precise
determination of the long range part of the PEC, for example from
photoassociation spectroscopy or from a molecular beam experiment.

The results of the LIF experiments, however, suggested that in
the case of Ca$_2$ it might be possible to make a reasonable
description of the long range part of PEC already by means of a
classical spectroscopy. Here we mean the shallow potential well
of the Ca dimer, which lead to thermal
population even of the highest rovibrational levels for normal
working temperatures (1220 K), so that a single laser scheme
could be applied for their excitation.
Of great importance is also the possibility to extend the
description of the PEC by a long range dispersive expansion already
from 9.4 \AA\ \cite{Allard:02}.

For detecting transitions from ground state levels with $v''>35$
we applied the Filtered Laser Excitation (FLE) technique, which
was used also by Hofmann {\it  et al.} \cite{Hofmann:86} for
studying the A$^1\Sigma_{\mathrm u}^{+}$ excited state in Ca$_2$.
With this technique we collected spectroscopic information for
about 24 highly excited levels of the X state with $v''$ up to
38. The outer classical turning point of the last observed level
with $v''=38$ and $J''=10$ is at 20 \AA\ and is located on the
rotationally reduced potential only 0.2 \rcm\ below the molecular
asymptote.

Although the FLE resolution was limited by the Doppler broadening, we
will demonstrate that combining the data from \cite{Allard:02} and
this experiment, it is possible to reduce the
uncertainty of the experimentally determined long range coefficients, and
especially C$_6$ to an extent, to compare them with the results
of the most recent theoretical predictions \cite{Porsev:02,Moszynski:01}.

In section~\ref{exper} we explain the experimental set up and
discuss the achieved experimental resolution and
uncertainties. Section~\ref{PEC}  summarizes the methods for
constructing of the PEC. Analysis of the experimental data are
performed in section~\ref{uncer}, where we employed a Monte Carlo
simulation in order to determine the error limits of the fitted long
range parameters of the PEC.

\section{Experiment}
 \label{exper}

The calcium dimers are obtained in a heat pipe oven already
described in Ref. \cite{Allard:02}. The experiments are performed
at helium or argon buffer gas under a typical pressure of $20-50$
mbar and an oven temperature of 1220 K. Working with argon as
buffer gas enables us to reduce the pressure to a lower value
than with helium without reducing the life time of the heat pipe
due to solid calcium closing the optical path. In the frequency
range accessible by the lasers in our group the most favorable
Franck-Condon factors between the levels close to the asymptote
of the X$^1\Sigma^{+}_{\mathrm g}$ state and the levels of the
B$^1\Sigma_{\mathrm u}^+$ state ($v^{\prime}=2$) are less than
$10^{-2}$. In order to achieve a sufficient signal-to-noise ratio
using Fourier transform spectroscopy a long term stability in the
operation of the oven is needed which is not achievable with our
present apparatus. In addition, as it was mentioned in
Ref.\cite{Allard:02}, the discrete fluorescence is accompanied by
a strong background emission due to bound-bound-free and
free-bound-free molecular transitions. To overcome this problems
we chose the filtered laser excitation (FLE) technique for direct
excitation of the levels close to the asymptote of the ground
state.

\begin{figure}
  \centering
\epsfig{file=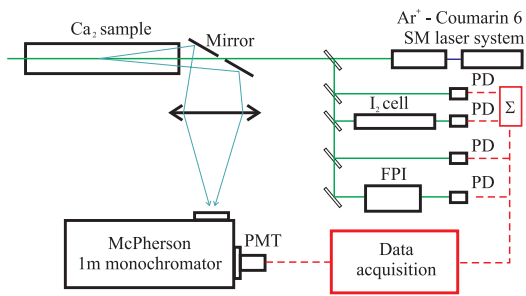,width=0.8\linewidth}
  \caption{Experimental setup}\label{Setup}
\end{figure}

A 1 m monochromator (GCA/McPherson Instruments) used as a narrow
band pass filter of a typical window width 2 cm$^{-1}$ is set on
a strong line of the fluorescence progression when exciting a
transition to a selected rotational level with $v^{\prime}=2$ of
the B state using a single mode dye laser (Coumarine 6). Then the
laser operating around 550~nm (60 mW on the Ca$_2$ sample) scans
the frequencies of transitions from near asymptote levels of the
ground state to the selected $v^{\prime}=2$ level. Laser
frequencies resonant with such transitions will result in
fluorescence, which matches the transmission band of the filter
and can be detected. The signal after the monochromator is
recorded as a function of the laser wavelength by a
photomultiplier (Hamamatsu, R928). Although many other
transitions of the X-B system contribute to the absorption
spectrum of Ca$_2$ in the scanned spectral region, only
excitations which decay into the selected frequency window will be
registered. Thus this selective technique provides greatly
simplified absorption spectra in a region where the weak
transitions from the asymptotic levels of interest are
practically completely overlapped by other much stronger
transitions.

\begin{figure}
  \centering
\epsfig{file=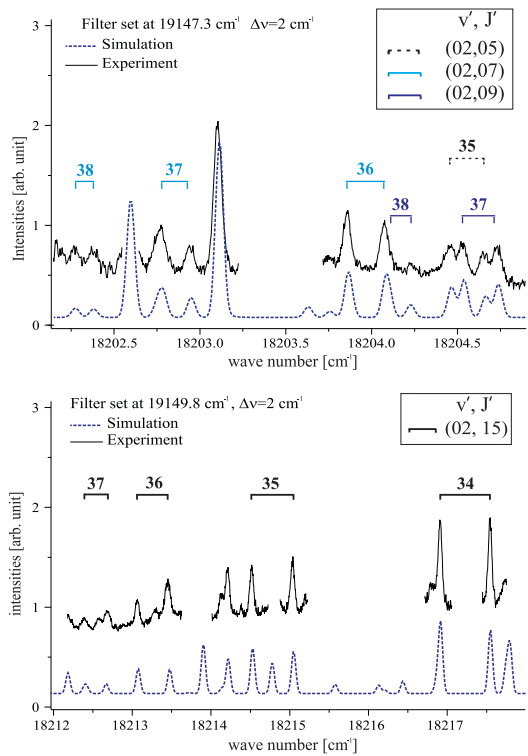,width=0.8\linewidth}
  \caption{Portions of filtered laser excitation spectra obtained with
   different detection windows. As dotted lines the synthetic spectra
  used for assignment of the observed spectral features are shown.}\label{Spectra}
\end{figure}

The relative positions of the FLE lines are calibrated using the
frequency comb of a temperature stabilized Fabry-Perot
interferometer with a free spectral range of 149.75(1) MHz. The
absolute line position is obtained using the absorption spectrum
of iodine vapor at room temperature (fig.~\ref{Setup}) calibrated
with the IodineSpec calculating software with which the positions
of the iodine lines are predicted with an accuracy better than
3MHz in the studied frequency range
\cite{toptica:02,KandTpriv:02}. In order to improve the
signal-to-noise ratio each spectrum is recorded several times.
Before the averaging procedure, the recorded signal is normalized
to the laser intensity to minimize the influence of its
variations with frequency and time. The typical width of the
observed lines is 0.05~cm$^{-1}$ corresponding to the Doppler
broadening. The intensity of the slowly varying background due to
free-bound-free or free-bound-bound transitions of the B-X system
is comparable to the intensity of the strongest lines and is
mainly responsible for the noise in the spectra. Taking account
of the linewidth, the signal-to-noise ratio and the uncertainty
in the calibration with the I$_2$ lines, the uncertainty of the
absolute frequency for the strongest lines is estimated to be
0.0035~cm$^{-1}$. Two typical spectra are shown in
fig.\ref{Spectra} for low and intermediate $J$. The experimental
traces as solid lines contain some gaps which were not scanned.
The number above the line gives $v''$ and the inset gives $v'$
and $J'$ of the excited state.

\begin{figure}
  \centering
\epsfig{file=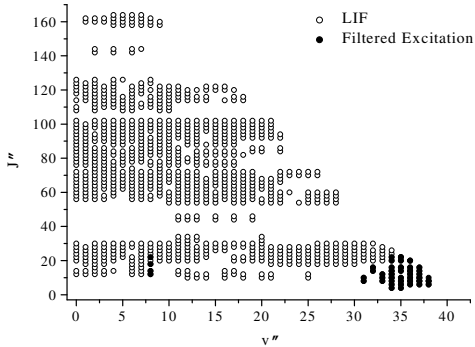,width=0.8\linewidth}
  \caption{The range of $v''$ and $J''$ quantum numbers of the observed
levels in the X$^1\Sigma^{+}_{\mathrm g}$ state of $^{40}$Ca$_2$.}
 \label{datafield}
\end{figure}

Using the FLE technique we observed transitions from 44 ground
state levels and among them 24 close to the asymptote with
$v''\ge 35$. The last observed levels are ($v''=38$, $J''=6, 8,
10$) as summarized in the data field in fig. \ref{datafield}. The
assignment of the new observed lines is performed using the
potential energy curve determined in our previous study of the
ground state \cite{Allard:02} and a numerical potential of the B
state determined from the data available in the literature
\cite{Balfour:75}. Simulations of the transitions by these
potentials for selected detection windows are presented in figure
\ref{Spectra} as dotted lines. These lines are shifted vertically
with respect to the observed traces in order to show clearly the
quality of the simulation. The intensity of the synthetic
spectrum is determined by the appropriate Franck-Condon factors,
by the thermal population of the ground state levels and by the
selected detection window. The small differences between the
observed and the simulated spectral features could be attributed
to the limited accuracy of the B state PEC (0.05 \rcm) and to the
fact that the used ground state potential was determined for
levels with only $v'' \le 35$. Also, only transitions in the main
calcium isotopomer, $^{40}$Ca$_2$ were simulated. The unambiguity
of the assignment was usually proved further by the observation
of relatively long (5-6 vibrational quanta) and self consistent
vibrational progressions of P,R doublets.

With the present set-up we can study the influence of the
collisions between calcium dimers and the buffer gas by
changing its pressure since no additional instrumental broadening
on the lines is present. We found that the collisional broadening
and the shift of the observed lines are much smaller than what was
previously assumed \cite{Allard:02}. This means that the
experimental uncertainties of the transitions obtained in that study
were overestimated.

For completeness we enriched our data with progressions
obtained by LIF using the 501~nm, 488~nm and 457 nm lines of an Argon
ion laser (Spectra Physics, BeamLok 2060) recorded by Fourier transform spectroscopy.
Most of the transitions which are known to be excited by these lines
\cite{Vidal:79} were observed in this study.
The present set of transitions representing all the LIF and FLE
data covers a range of rotational quantum numbers from $J''= 4$
to $164$ and a range of vibrational quantum numbers from $v''= 0$
to $38$. We have recorded a total of 3580 transitions resulting 
in 924 levels of the ground state \footnote{See EPAPS Document No. 
... for the list of the transition frequencies.
This document may be retrieved via the EPAPS 
homepage (http://www.aip.org/pubservs/epaps.html) or
from ftp.aip.org in the directory /epaps/. 
See the EPAPS homepage for more information}.

During this study six atomic calcium lines were also observed
around $1.9 \mu$m, which correspond to the
$^{3}$D$_{1,2,3}~\rightarrow~^{3}$P$_{1,2,3}$ transitions, when
irradiating the sample by Ar$^+$ laser lines. These transitions
have been already observed in Mg in a similar experiment
\cite{Scheingraber:77}. Here they could be attributed to
predissociation of the molecular B state through a coupling with
repulsive states correlated to the $^{3}$D + $^{1}$S asymptotes.
It is worth mentioning that the
$^{3}$D$_{1,2,3}~\rightarrow~^{3}$P$_{1,2,3}$ transitions were
observed also at temperatures down to 950~K, which suggests that
the first step of the process is predominantly excitation of
pairs of free Ca atoms by the laser radiation to the molecular B
state. The intensity ratio between the infrared emission and the
laser light was highest for the 457 nm Ar$^+$ line and decreases
with the increase of the laser wavelength.

After a new examination of the previous Fourier transform spectra using
the new potential energy curve which will be presented in the following,
three unassigned weak molecular progressions were found to belong
to the calcium isotopomer $^{44}$Ca$^{40}$Ca. The exciting transitions
were identified as (0, 42)$\leftarrow$(4, 43) and
(0, 72)$\leftarrow$(5, 73) at 18788.36 cm$^{-1}$ and
(0, 111)$\leftarrow$(8, 112) at 18787.36 cm$^{-1}$.

\section{Construction of PEC}
\label{PEC}

Following the approach adopted in the previous report
\cite{Allard:02}, information which concerns only the ground
state was extracted from the raw spectroscopic data. From
transition frequencies with a common upper state level,
differences between ground state levels were calculated. The PEC
for the ground state was then constructed, which describes
these experimentally observed differences. Similar to
Ref.\cite{Allard:02} the possible combinations of line
frequencies within a progression were restricted only to pairs of
one P and one R lines, forming a total of 8500 differences.

The methods for construction of PEC were discussed in detail in
Ref. \cite{Allard:02}  and will be summarized only briefly here.

The first method represents the potential as a truncated expansion
over analytic functions:

\begin{equation}\label{anform}
  U(R)=\sum_{i=0}^{n}a_i\left(\frac{R-R_m}{R+bR_m}\right)^i,
\end{equation}

\noindent where $a_i$, $b$ and $R_m$ are parameters
($R_m$ is close to the equilibrium distance).
For short and long internuclear distances the
potential is smoothly extended respectively for $R \leq R_{\mathrm {inn}}$
with

\begin{equation}
A+B/R^{12}
\end{equation}

\noindent and for $R \geq R_{\mathrm {out}}$ with

\begin{equation}
D_{\mathrm e}-C_6/R^6-C_8/R^8-C_{10}/R^{10} \mbox{.}
\label{LR}
\end{equation}

Here $D_{\mathrm e}$ is the value of the dissociation energy
defined with respect to the minimum of the PEC. Since $R_{\mathrm
{out}}$ is usually chosen within the region where the
contribution of the exchange energy is negligible, no additional
dumping (or cut off) functions for the dispersion coefficients
are introduced.

Parameters $a_i$
and $C_6$ and $C_8$ are fitted in a nonlinear
fitting procedure (for details see  \cite{Samuelis:00}), while
$A$, $B$, $C_{10}$ and $D_{\mathrm e}$ are adjusted
by the program in order to ensure smooth connection between the
extensions and the analytic form. $R_m$, $b$ and the connecting
points $R_{\mathrm {inn}}$ and $R_{\mathrm {out}}$ are kept
fixed to values, which allow fast convergence of the fitting
routine.

The second method defines the PEC for short and intermediate
internuclear distances as a set of points connected with cubic
spline function. For large internuclear distances the same long
range expansion as in (\ref{LR}) is used. The values of the
potential $U_i$ in a preselected grid of points,
 $D_{\mathrm e}$, $C_6$, $C_8$ and $C_{10}$ are treated as
fitting parameters. They are determined in an iterative fitting
routine, which linearizes the problem by using a modified
version of the inverted perturbation approach
\cite{ipaasen}. The connecting point $R_{\mathrm{out}}$ is chosen
in the following way. Initially, the PEC is constructed in a pointwise
form up to 13 \AA.  Then only the long range parameters are
fitted assuming that $R_{\mathrm  {out}}=9.5$ \AA. Although both
sections of the PEC are determined independently, their shapes
turn out to be almost identical between 9.5~\AA\ and 13 \AA. As
a last step the $C_{10}$ coefficient of the long range expansion
is slightly adjusted (the change does not exceed few percent)
in order to fit best the shape of the pointwise potential
between 9.4 \AA\ and 10 \AA. The crossing point of the pointwise
and the long range curves is taken as $R_{\mathrm{out}}$. The new value of
$R_{\mathrm{out}}$ differs from that which was used when fitting
the long range parameters. Usually the change is small (typically
0.1 \AA) and since the two sections are very close in this
region, the effect of such a change on the quality of the fit is
negligible. Of course, a second iteration could be performed with
the new value of $R_{\mathrm{out}}$, but usually it is necessary
only in the beginning, when the initial values of the fitting
parameters are far from the best ones.

Before fitting the PEC to the new data set, the experimental
errors of the data obtained by LIF were reanalyzed, since the
value of the normalized standard deviation $\sigma \approx 0.45$ given in
the previous study (Ref.~\cite{Allard:02}) signals their
overestimation. The value of the uncertainty before was chosen
to be 0.01 \rcm\ for a strong LIF line in order to take into account
possible frequency shifts due to collisions. Since the new FLE
measurements showed that the probable influence of the
temperature and the buffer gas pressure are much smaller than
expected we reduced the errors of all LIF frequencies by a factor
of 0.6, which will give a more realistic estimate on the error
limits of the fitted PEC parameters.

\begin{table}
  \centering
  \caption{Parameters of the analytic representation of the X$^1\Sigma^{+}_{\mathrm
  g}$  state potential energy curve in $^{40}$Ca$_2$.}
 \label{anpottab}
\begin{tabular*}{0.9\linewidth}{@{\extracolsep{\fill}}lr}\hline
   \multicolumn{2}{c}{$R \leq$ 3.66 \AA}    \\
   $R_\mathrm{inn}$ & 3.66 \AA    \\
   $A$ & $-2.9714\times 10^{2}$  \rcm \\
   $B$ & $7.209\times 10^{9}$ \rcm\AA$^{12}$ \\
   &                                   \\
   \multicolumn{2}{c}{3.66 \AA\ $< R <$ 9.5 \AA}    \\
    $b$ &   -0.5929              \\
    $R_\mathrm{m}$ & 4.277277 \AA               \\
    $a_{0}$ & $ 0.001287$                 \rcm\\
    $a_{1}$ & $-2.57153863528197002$ \rcm\\
    $a_{2}$ & $ 3.79611687289805877\times 10^{3}$ \rcm\\
    $a_{3}$ & $ 3.82947943867555637\times 10^{2}$ \rcm\\
    $a_{4}$ & $-2.74470356912936631\times 10^{3}$ \rcm\\
    $a_{5}$ & $-3.23378807398046092\times 10^{3}$ \rcm\\
    $a_{6}$ & $ 3.70205119299758223\times 10^{2}$ \rcm\\
    $a_{7}$ & $ 6.35318559107446436\times 10^{3}$ \rcm\\
    $a_{8}$ & $-7.39783474312859562\times 10^{3}$ \rcm\\
    $a_{9}$ & $-1.90759867971015337\times 10^{4}$ \rcm\\
   $a_{10}$ & $ 5.41779135173975228\times 10^{4}$ \rcm\\
   $a_{11}$ & $ 4.40527349765557083\times 10^{4}$ \rcm\\
   $a_{12}$ & $-1.55406021572582802\times 10^{5}$ \rcm\\
   $a_{13}$ & $-8.35826911941128783\times 10^{4}$ \rcm\\
   $a_{14}$ & $ 2.13873243831604603\times 10^{5}$ \rcm\\
   $a_{15}$ & $ 1.56022970979522303\times 10^{5}$ \rcm\\
   $a_{16}$ & $-1.56329579530082468\times 10^{5}$ \rcm\\
   $a_{17}$ & $-1.46822446075956163\times 10^{5}$ \rcm\\
   $a_{18}$ & $ 2.74480910039127666\times 10^{4}$ \rcm\\
   $a_{19}$ & $ 7.11882274192053592\times 10^{4}$ \rcm\\
   $a_{20}$ & $-7.63044568335207146\times 10^{2}$ \rcm\\
   &                                   \\
   \multicolumn{2}{c}{$R \geq 9.5$ \AA\ }\\
   $R_\mathrm{out}$ & 9.5 \AA               \\
  $D_{\mathrm e}$ & 1102.076 \rcm             \\
  $C_6$ &  $1.0030\times 10^{7}$ \rcm\AA$^6$      \\
 $C_{8}$ & $ 3.87\times 10^{8}$ \rcm\AA$^8$   \\
 $C_{10}$ & $ 4.39\times 10^{9}$ \rcm\AA$^{10}$\\
 \\
\multicolumn{2}{c}{Additional parameter}\\
\\
 $D_{0}=1069.871$ \rcm   &          \\
\hline

\end{tabular*}

\end{table}

\begin{table}
  \centering
  \caption{Parameters of the numeric representation of the
  X$^1\Sigma^{+}_{\mathrm g}$ state potential energy curve in $^{40}$Ca$_2$.}
 \label{ipatab}
 \begin{tabular*}{0.9\linewidth}{@{\extracolsep{\fill}}rrrr}\hline
 R [\AA] & U [cm$^{-1}$]& R [\AA] & U [cm$^{-1}$]\\ \hline
 3.096980 & 9246.6895 &      5.678571 &   636.3741  \\
 3.188725 & 6566.7325 &      5.809524 &   684.9589  \\
 3.280470 & 4525.7282 &      5.940476 &   728.9235  \\
 3.372215 & 3090.9557 &      6.071429 &   768.5976  \\
 3.463960 & 2134.2175 &      6.202381 &   804.2551  \\
 3.555705 & 1475.2425 &      6.333333 &   836.2419  \\
 3.647450 & 1004.5043 &      6.464286 &   864.8746  \\
 3.739195 &  661.4123 &      6.595238 &   890.4666    \\
 3.830940 &  410.6117 &      6.726191 &   913.2923   \\
 3.922685 &  234.0001 &      6.857143 &   933.6417   \\
 4.014430 &  116.0996 &      6.988095 &   951.7718   \\
 4.106174 &   44.5437 &      7.119048 &   967.8632   \\
 4.197920 &    8.6885 &      7.250000 &   982.2159   \\
 4.289664 &    0.1760 &      7.500000 &  1005.2497   \\
 4.381409 &   11.9571 &      7.750000 &  1023.6698   \\
 4.500000 &   48.5948 &      8.000000 &  1038.3262    \\
 4.630952 &  106.9081 &      8.358974 &  1054.3861     \\
 4.761905 &  175.7311 &      8.717949 &  1066.0579     \\
 4.892857 &  248.8199 &      9.076923 &  1074.5969     \\
 5.023809 &  322.3873 &      9.435897 &  1080.8961     \\
 5.154762 &  393.7222 &      9.794872 &  1085.5974     \\
 5.285714 &  461.4555 &     10.303419 &  1090.2990     \\
 5.416667 &  524.6311 &     10.811966 &  1093.5160     \\
 5.547619 &  582.9870 &     11.611111 &  1096.6870     \\
   &&&\\
 \multicolumn{2}{l}{$D_{\mathrm e}=1102.060$ \rcm} & & \\
 \multicolumn{2}{l}{$R_{\mathrm{out}}=9.44$ \AA} &
 \multicolumn{2}{l}{$C_8$=3.808$\times 10^{8}$ \rcm\AA$^{8}$} \\
 \multicolumn{2}{l}{$C_6$=1.0023 $\times 10^{7}$ \rcm\AA$^{6}$} &
 \multicolumn{2}{l}{$C_{10}$=5.06$\cdot 10^{9}$ \rcm\AA$^{10}$}\\
\\
 \multicolumn{2}{l}{Additional parameter}  & \\
\\
 \multicolumn{2}{l}{$D_0 =1069.868$ \rcm}\\

\\
 \hline
\end{tabular*}

\end{table}

The analytic potential listed in table~\ref{anpottab}
describes the differences between the
observed spectral lines from both experiments with a standard
deviation $\sigma=0.0064$ \rcm\ and normalized standard deviation
$\bar{\sigma}$=0.69.
The quality of the PEC for the near asymptotic levels
is estimated by calculation the standard deviation with a reduced
set of differences, where for each difference at least one level
belongs to $v'' \ge 35$. We
obtained $\sigma_{{35}}=0.0092$ \rcm\ and $\bar{\sigma}_{{35}}=0.92$.
The same parameters for the numerical potential (table~\ref{ipatab}) are
$\sigma=0.0068$ \rcm, $\bar{\sigma}=0.74$ and
$\sigma_{{35}}=0.0092$ \rcm, $\bar{\sigma}_{{35}}=0.89$.
In order to calculate the value of the pointwise potential in the
range $R<9.44$ \AA, a natural cubic spline
through all the grid points should be used.
The parameters of both curves are chosen such, that their minima
are set to zero. In order to facilitate the comparison between the dissociation
energies $D_{0}$ defined with respect to the lowest rovibrational
level ($v''=0, J''=0$), their values are given in tables~\ref{anpottab} and \ref{ipatab}
as additional parameters.

\section{Uncertainties of the long range parameters}
\label{uncer}

The motivation of this study is not
only to describe the rovibrational structure of the $^{40}$Ca$_2$ ground
state. We rather want to examine to which extent the performed
spectroscopic study is able to go beyond the reproduction of the
observed differences between ground state levels. Of greatest
interest is the reliability of the determined long
range parameters (and especially $C_6$) since they play an
important role in phenomena involving two interacting
cold Ca atoms in the ground state.

Our analysis relies on two main assumptions:

\begin{itemize}

 \item the interaction between two Ca atoms can be described within an
 adiabatic picture, through a single channel model applying
 the one dimensional \Sch\ equation with an effective potential energy curve;

  \item the shape of the potential for internuclear distances $R\ge 9.4$~\AA\
 can be described with the long range expression (\ref{LR}), neglecting the
  exchange energy.

 \end{itemize}

Due to the zero net electronic and nuclear spin in the case of $^{40}$Ca
and the large energy separation between the ground state and the lowest
excited states the first assumption seems to be well justified. The second
one is supported by the expected small value of the exchange energy
\cite{Radzig,Allard:02} at 9.5 \AA\ and the dominant van der
Waals character of the interaction.

Under these assumptions we ask the question: what are the
variations of the PEC parameters around the fitted ones
which are still in agreement with the experimental data?

In Ref.~\cite{Allard:02} the answer was given by plotting a
contour plot of the likehood function $\chi^2$ by varying the long
range parameters. For the nonlinear fitting procedures
it is, however, not straightforward how to use the
contour plot in order to determine the confidence interval of the
parameters since it is not obvious to which confidence limit a
given contour corresponds. That is why there is some ambiguity in
the probable errors of the fitted parameters determined only by
using contour plots. The use of the matrix of variances and
covariances, which is obtained by the fitting routines in the linearized
form is not useful in our case. The reason is that even if we assume the
errors of the experimental frequencies being independent and
following a normal distribution, the quantities which enter the fit are
actually differences between these frequencies. Consequently
their errors are not independent and we are not allowed to
interpret the above mentioned matrix as in the case of
independent data points.

A possible way to get the probability distribution of the fitted
parameters is to perform a Monte Carlo simulation of synthetic data.

\subsection{Monte Carlo simulation}

Since our experimental data come with some measurement uncertainties,
which we assume to be normally distributed, the set $c^{(0)}$ of the
fitted parameters of the PEC will differ from the true one $c^{(t)}$.
If we perform
a series of similar experiments the experimental data sets will be
slightly different and consequently we will obtain different sets
of fitted parameters $c_{exp}^{(i)}$. Having a sufficient number of
measurement sets we could plot the distribution of $c_{exp}^{(i)}$ and then
decide what are the most probable values and what are their
uncertainties.

\begin{table}
  \centering
  \caption{Parameters of the long range expansion for the
X$^1\Sigma^{+}_{\mathrm g}$ state in $^{40}$Ca$_2$ derived in
this study and compared with the most recent data from the
literature.}

  \begin{tabular*}{0.9\linewidth}{@{\extracolsep{\fill}}lrr}

 & This study & Other sources \\  \hline
 $D_{\mathrm e}$, \rcm &&1095.0(5) \cite{Vidal:79} \\
                                   &&1102.08(9) \cite{Allard:02}\\
 $D_0$, \rcm &1069.868(10)&1069.88(9) \cite{Allard:02}\\
 $C_6 \times 10^{7}$ \rcm\AA$^6$ & 1.003(33) & 1.070(6) \cite{Porsev:02}\\
                             &&1.098 \cite{Moszynski:01}\\
                             &&1.15 \cite{Vidal:79}\\
                             && $1.02 \div 1.12$ \cite{Allard:02}\\
 $C_8\times 10^{8}$ \rcm\AA$^8$  &$3.15\div 4.46 $ & 3.27 \cite{Moszynski:01}\\
                                  &  & $1.1 \div 3.8$ \cite{Allard:02}\\
 $C_{10}\times 10^{9}$ \rcm\AA$^{10}$ & $1.7 \div 8.4$ & 4.74 \cite{Moszynski:01}\\
                                      &&$3.7 \div 17.0$
                                      \cite{Allard:02}\\\hline
 \end{tabular*}
 \label{param} \end{table}

So let us assume that the fitted potential $U^{(0)}(R)$, described with
a set of parameters $c^{(0)}$, is not too far from the true
one. Then, knowing the experimental uncertainty associated with
each observed transition frequency, we can synthesize a
set of experimental data which will have exactly the same
structure as the real one (measurement errors, distribution of
$v$ and $J$ quantum numbers). For this goal transition frequencies
are formed
using the eigenvalues for $U^{(0)}(R)$ (since the upper state
levels play no role in our analysis their energies could be set
to zero) and adding to each calculated frequency a small random
quantity normally distributed with mean value zero and a standard
deviation equal to the experimental error. Then we transform the synthetic
frequencies into synthetic differences and use the same fitting
procedure in order to obtain a new set of parameters $c_{sim}^{(i)}$.
Performing a large number of simulations we can plot the
distribution of $c_{sim}^{(i)} - c^{(0)}$.  Since the simulated data
are related statistically to $U^{(0)}(R)$ in the same way as the
real experimental data to the true potential and since these two
potentials were assumed to be close to each other the obtained
distribution should be not too different from that of $c_{exp}^{(i)} -
c^{(t)}$.

\begin{figure*}
  \centering
\epsfig{file=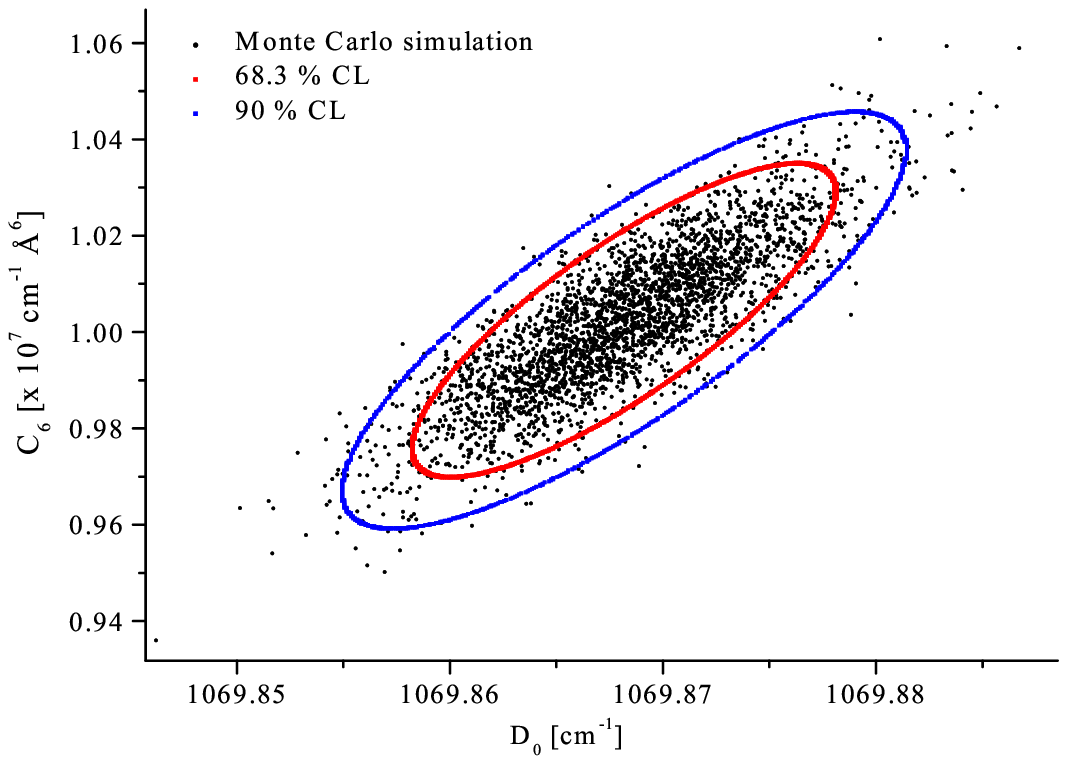,width=0.48\linewidth}
\epsfig{file=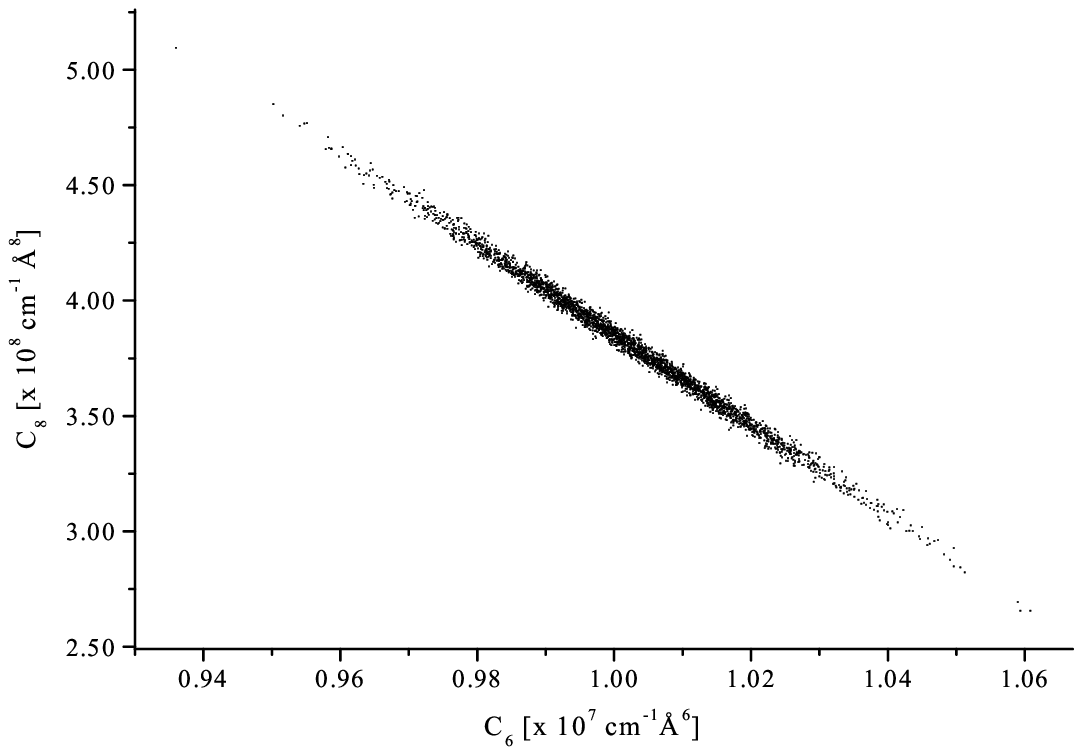,width=0.48\linewidth}
  \caption{Projections of the long range
parameters distribution realized with 3000 simulated points
on the $D_0$-$C_6$ and $C_6$-$C_8$ planes. CL stays for
confidence limit.}\label{D0distr}
\end{figure*}

In order to realize these ideas we used the pointwise
representation of the potential since in the present realization of
our computer codes it allows more flexibility in connecting the
long range expansion with the potential curve at intermediate
internuclear distances. The prize to pay for this is that the
first derivative of the PEC might have a small ``kink'' in
$R_{\mathrm{out}}$. This problem, however, does not influence
the analysis of the PEC and it could be avoided by slight
adjustment of the pointwise curve. What
we win is that all the long range parameters are fitted to the
experimental data independently of the actual shape of the
potential at intermediate internuclear distances. The
continuity of the potential is ensured mainly by varying the
connecting point $R_{\mathrm{out}}$ and to a smaller extend
$C_{10}$. Analysis similar to those, performed in Ref.~\cite{Allard:02}
showed that the present body of experimental data fixes the inner part
of the PEC approximately between 3.5 \AA\ and 9.5 \AA. Possible variations
in this region are
connected mainly with the experimental uncertainties and the influence
of the selected representation for the potential is negligible.
Therefore we may expect that the choice of the connecting point $R_{\mathrm{out}}$
around 9.5 \AA\ makes our analysis almost independent of the functions used to
model the PEC at short internuclear distances.

Having generated a set of synthetic differences
$\Delta E^{(i)}_{v_1J_1,v_2J_2}= E^{(i)}_{v_1J_1} - E^{(i)}_{v_2J_2}$
we treat them as the experimental data.
$U^{(0)}(R)$ is taken as initial potential and a small correction
to it is fitted which should minimize the difference between
$\Delta E^{(i)}_{v_1J_1,v_2J_2}$ and the differences calculated
with the corrected potential. The correction is defined as
follows:

\begin{equation}
\delta U(R)=\sum_{i=1}^N \delta u_i S_i(R) \mbox{,   for R $<$ R$_{out}$}
\end{equation}

\noindent and

 \begin{equation}
 \delta U(R)=\delta D_e-\frac{\delta C_6}{R^6}-\frac{\delta C_8}{R^8}-\frac{\delta C_{10}}{R^{10}}
 \mbox{,   for R $\ge$ R$_{out}$,}
 \end{equation}

\noindent where $S_i(R)$ are defining functions of the pointwise potential
(see \cite{MyPhD}),
$\delta u_i$ are the values of the correction in an equidistant grid of
internuclear distances $R_i$ and $\delta D_e$, $\delta C_6$, $\delta C_8$,
$\delta C_{10}$ are the corrections to
the long range parameters. Using the theory of perturbations the
shift of the  difference $\Delta E^{(i)}_{v_1J_1,v_2J_2}$ due to
$\delta U(R)$ can be written as:

\begin{equation}
\begin{split}
\delta(\Delta E^{(i)}_{v_1J_1,v_2J_2})=&\sum_{i=1}^N
\delta u_i(K_i^{v_1J_1}-K_i^{v_2J_2})\\+ \; &\delta
D_e(L_0^{v_1J_1}-L_0^{v_2J_2})\\ -&\sum_{j=6,8,10} \delta
C_j(L_j^{v_1J_1}-L_j^{v_2J_2}) \mbox{ ,} \\ \label{fiteq}
\end{split}
\end{equation}

\noindent where

\begin{equation}
K_i^{vJ}=\int_{0}^{R_{out}}\Psi^2_{vJ}(R)S_i(R)dR
\label{srcoef}
\end{equation}

\begin{equation}
L_j^{vJ}=\int_{R_{out}}^{\infty}\Psi^2_{vJ}(R)R^{-j}dR
\label{lrcoef}
\end{equation}

\noindent are the corresponding mean values of $S_i(R)$ and
$R^{-j}$ calculated with the wave functions $\Psi_{vJ}(R)$ of the
levels forming the difference.

Note that in principle the exact presentation of $U^{(0)}(R)$ is not
of importance since we study only the possible small variations around
it which are still allowed by the experimental data.
The advantage of expressing the correction to the potential in a
pointwise form (cubic spline function is used for interpolation)
is that the influence of the parameters $\delta u_i$ is localized in
$R$, since away from $R_i$ $S_i(R)$ decays exponentially \cite{ipaasen}.
This gives a high flexibility in constructing the needed form of the
correction. Additionally, the correlations between $\delta u_i$ are
introduced mainly by the experimental data and not by the
selected basis functions.

With each set of simulated data a new fit is performed which adjusts
$\delta u_i$, $\delta D_e$, $\delta C_6$, $\delta C_8$ and
$\delta C_{10}$. Along with this we examine
the contributions due to $\delta u_i$ and the long range corrections
to the fitted differences (the sums in Eq.~\ref{fiteq}).
As it could be expected, the changes of the PEC for $R<R_{out}$ are
very small ($\delta u_i \sim 0.01$ \rcm)
for all sets of synthetic data due
to the large amount of experimental observations in this region.
In addition the corrections to the differences due to $\delta u_i$,
i.e. the first term of the right side of eq.~\ref{fiteq},
turn out to be smaller than the corresponding experimental uncertainties
and in principle one could neglect them.
We checked this and, indeed, the distribution
of the long range parameters was
almost identical for simulations with and without fitting $\delta u_i$.
Since this is true for different numbers of $\delta u_i$ parameters, we
may consider it as a proof of the statement made above, that in the
present case the determination of the long range parameters is almost
independent on the model used to describe the inner part of the PEC.

The positions of the energy levels and the molecular asymptote
$D_e$ in Ref.~\cite{Allard:02} were defined with respect to the
minimum of the PEC, $U_{\mathrm{min}}$. Then we estimated that
the possible variation of $U_{\mathrm{min}}$ for different
representations of the potential is of the order of 0.01 \rcm,
which was smaller than the uncertainty of the dissociation
energy, 0.09 \rcm. With the extended experimental data set and
after the revision of the experimental uncertainties we are now
able to give a much better prediction of the dissociation energy
with uncertainty also of the order of 0.01 \rcm. Obviously
defining the dissociation energy with respect to
$U_{\mathrm{min}}$ will introduce an undesired increase of the
uncertainty for this long range parameter. Consequently, starting
from here we will consider only the value of the dissociation
energy $D_0$ defined with respect to the lowest rovibrational
level $E_{00}=E_{v''=0,J''=0}$. Indeed, fitting $\delta u_i$ to
the simulated data we convinced ourselves that although the shape
of the potential (and also $U_\mathrm{min}$) could vary slightly
from fit to fit, the overall effect on the position of $E_{00}$
is much smaller than the typical variation of $U_\mathrm{min}$.
Nevertheless, the contribution of $E_{00}$ to the uncertainty of
$D_0$ is taken into account.

In this way, the main steps of the performed Monte Carlo analysis are:

\begin{enumerate}
  \item Find the PEC $U^{(0)}(R)$ which can reproduce the experimental
  observations with smallest standard deviation;
  \item From the eigenvalues for $U^{(0)}(R)$ generate a set
  of synthetic data by adding normally distributed random deviations
  according the experimental uncertainties. This set has exactly
  the same structure as the experimental data;
  \item Perform a fit with the synthetic data as it was done with
  the original one and collect the obtained potential parameters in
  a distribution list;
  \item Repeat steps (2)-(4) until enough simulated long range
  parameters are collected in order to analyze their distribution.
\end{enumerate}

\subsection{Results}
\label{res}

In figure~\ref{D0distr} two projections of the long range
parameters distribution obtained from 3000 simulations
are shown. It is worth mentioning the strong correlation between
$C_6$ and $C_8$, which is the case also for the other pairs
of dispersive coefficients. In order to
determine the confidence regions for the parameters we construct
an ellipse in the four dimensional space of the parameters, whose
axes have such orientations and relative lengths that the
resulting surface follows the form of the distribution. Then we
change gradually the lengths of the axes by a constant factor
and count the number of points which are enclosed by the
ellipse. For example, the ellipse which contains 68.3~\% of the
points will determine the 68.3~\% (1$\sigma$) confidence region of the
parameters. The projections of the 68.3~\% and the 90~\%
confidence regions (strictly speaking the contours surrounding the
projections) are plotted in the left part of
figure~\ref{D0distr} for the pair of
parameters $D_0$ and $C_6$.

This allows us to determine the mean value and uncertainty of the dissociation
energy of the $^{40}$Ca$_2$ ground state with respect to the
lowest rovibrational level as $1069.868 \pm 0.010$ \rcm\ and of the $C_6$
coefficient as $(1.003 \pm 0.033)\cdot 10^{7}$ \rcm\AA$^6$ at
a confidence limit of 68.3~\%. The values and the uncertainties of
all the long range parameters are summarized in table~\ref{param}
and compared with the values from other sources.

\begin{figure}
  \centering
\epsfig{file=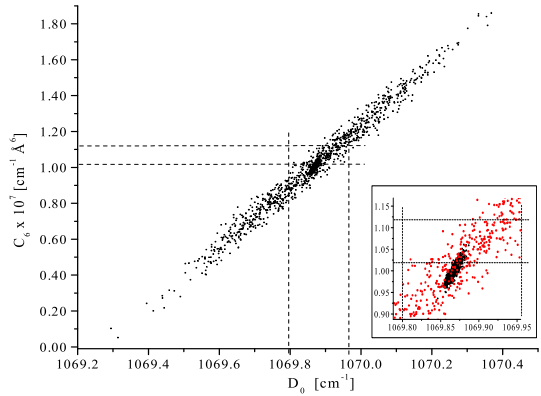,width=\linewidth}
  \caption{Projections of the long range
parameters distributions on the $D_0$-$C_6$ plane derived from
the LIF data from Ref.~\cite{Allard:02} and from this study (the black
spot in the centre). }\label{35-38}
\end{figure}

\begin{figure}
  \centering
\epsfig{file=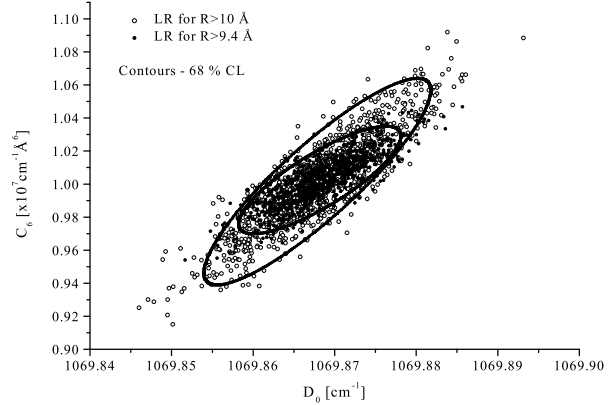,width=\linewidth}
  \caption{Projections of the long range
parameters distributions on the $D_0$-$C_6$ plane derived by
assuming the validity of the long range expansion from 9.4 \AA\
(solid circles) and from 10 \AA\ (open circles).}\label{10A}
\end{figure}

The highest observed rovibrational level in Ref.~\cite{Allard:02}
had $v''=35$ and it was concluded that with the available
experimental data set it is not possible to give any reasonable
estimation for $C_6$. Now, by using a Monte Carlo analysis we can
be more rigorous. In figure~\ref{35-38} we compare the
projections of two distributions on the $D_0$-$C_6$ plane.  The
first one is derived using only the data from
Ref.~\cite{Allard:02} before the revision of those experimental
errors. In order to obtain a meaningful estimation of the
dissociation energy of the X state, we restricted in
Ref.~\cite{Allard:02} the possible variation of C$_6$ within a
$\pm$ 5 \%\ interval around the theoretical value
\cite{Porsev:02}. This interval is denoted in fig.~\ref{35-38}
with horizontal dashed lines and leads to an uncertainty of the
dissociation energy similar to that, given in
Ref.~\cite{Allard:02}. After including the new LIF and FLE data
the distribution shrinks to the small black spot in the centre of
figure~\ref{35-38} (see also the inset), which is actually the
distribution from figure~\ref{D0distr}. It would be, however, not
correct to interpret the drastic reduction of the size of the
distribution only as a result of the transitions involving
rovibrational levels with $v'' > 35$. In fact the items, which
led to the distribution presented in figure~\ref{D0distr} are:

\begin{itemize}
  \item The assumption that the long range expansion (\ref{LR}) is
  valid starting from 9.4 \AA. This allows to link the values of the
  long range parameters with the frequency differences involving levels with
  $30 \le v'' \le 38$, i.e. the shape of the long range potential is roughly
  tested from  9.4 \AA\ up to $\approx 20$ \AA\
 (the classical turning point of the last observed
  level). A Monte Carlo simulation shows (figure~\ref{10A}) the
  change of the distribution when shifting $R_{\mathrm{out}}$ from 9.4 \AA\ only
  by 0.6 \AA\ to $R_{\mathrm{out}}=10$ \AA. Note, that in this case the loss
  of accuracy on $C_6$ is larger by a factor of two compared
  to the increase of the $D_0$ uncertainty;

  \item Availability of highly excited levels with $v'' = 38$, since they
  limit the variation of the dissociation energy;

  \item The enlargement of the experimental data, which
  allow accurate determination of the PEC for short
  and intermediate internuclear distances. Because of the large
  body of data, the long range parameters are almost independent
  of the exact representation of the rest part of the PEC,
  which is clearly indicated by the small variations of $\delta u_i$
  (eq.~\ref{fiteq}) during the statistical analysis;

  \item The corrected estimate of the experimental uncertainties
  of the data in Ref.~\cite{Allard:02}.

\end{itemize}

\subsection{The value of the $C_6$ coefficient}

The confidence interval for $C_6$ given in this paper $(1.003 \pm
0.033)\cdot 10^{7}$ \rcm\AA$^6$ does not overlap with the most recent
theoretical prediction of $(1.070 \pm 0.006) \cdot 10^{7}$ \rcm\AA$^6$
within the error stated in Ref.~\cite{Porsev:02}.

From our side, we realize that an important point in our analysis
is the assumption for the validity of the long range expansion
(\ref{LR}) down to $\approx 9.4$ \AA. As it was shown in
fig.~\ref{10A}, shifting $R_{\mathrm{out}}$ to larger internuclear
distances will increase the uncertainty on the fitted parameters.
The same will happen if we introduce additional terms into the
long range model (for example the exchange energy). Our analysis
can readily include these additional parameters, provided there
are new experimental observations which require this.

\subsection{The s-wave scattering length}

\begin{figure*}
  \centering
\epsfig{file=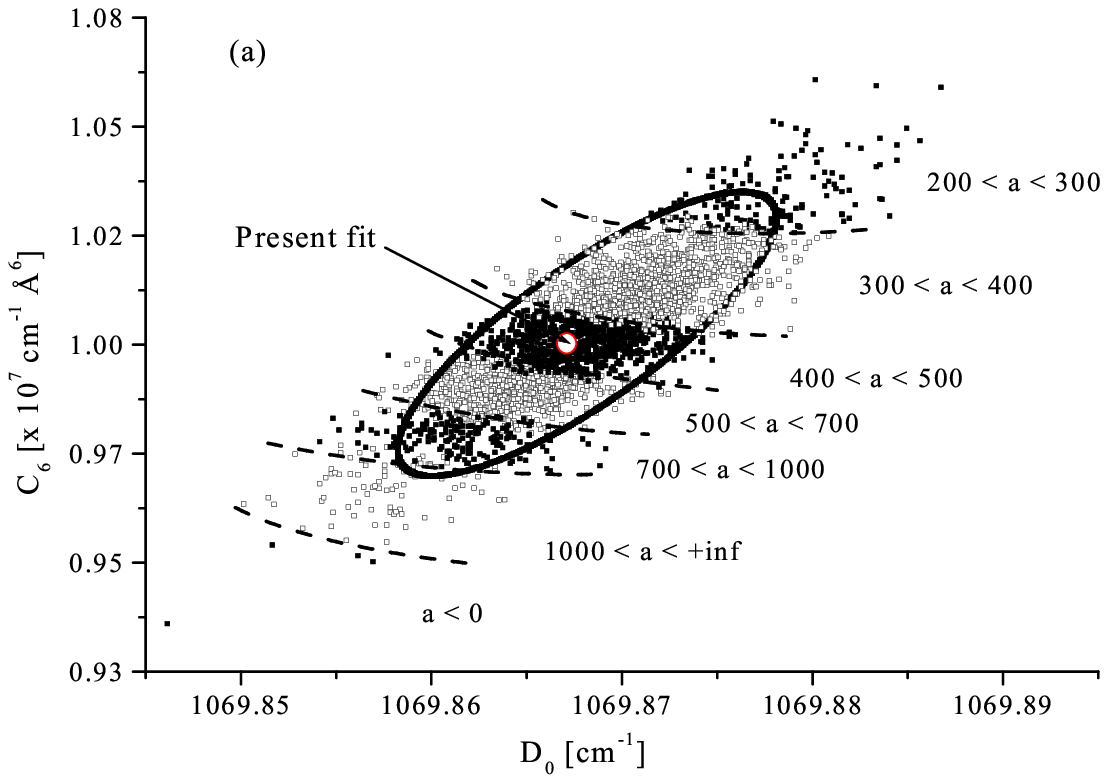,width=0.49\linewidth}
\epsfig{file=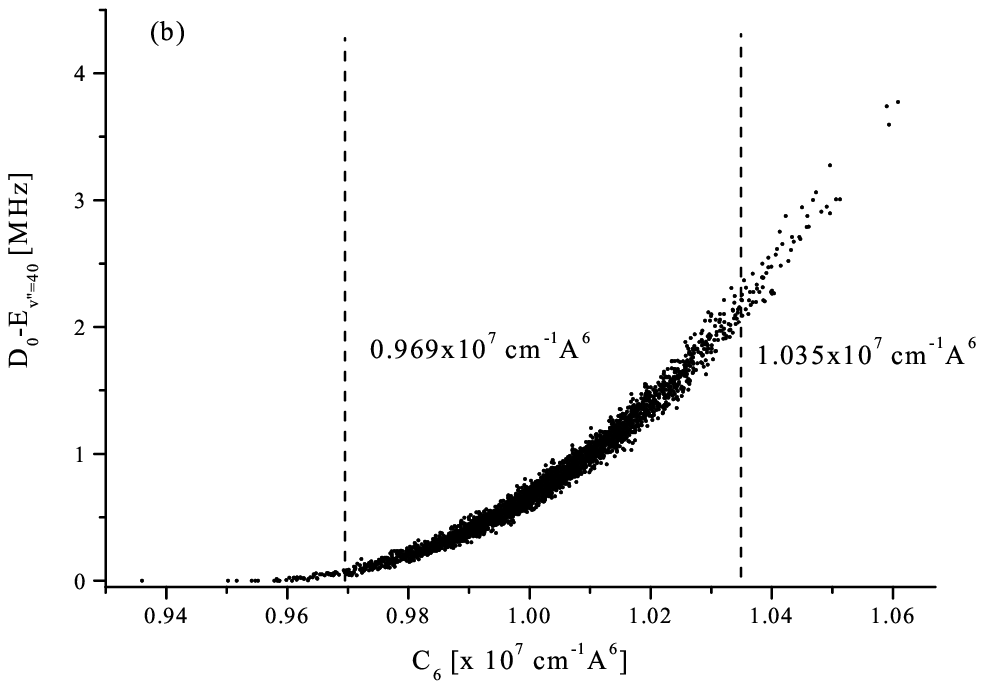,width=0.49\linewidth}
  \caption{Distribution of the s-wave scattering length  (a)
and the position of the last bound level with respect to the
asymptote (b) for the long range potentials obtained from the
Monte Carlo simulation.}\label{scat} \end{figure*}

In order to obtain the s-wave scattering length and its confidence
interval, it is straightforward to calculate its value
for each of the simulated PEC. These
calculations will show how reliably the spectroscopically
determined potential up to $\approx 20$ \AA\ can be applied to
model cold collisions. The results are presented in fig.~\ref{scat}a.
The position of the last bound level ($v''=40$) with respect to the
molecular assymptote, $D_0-E_{v''=40,J''=0}$ is shown in fig.~\ref{scat}b.
The values of the scattering length $a$ within the 68.3 \%\ confidence region
vary from $250
a_0$ to $1000 a_0$ (the Bohr radius $a_0\approx 0.52918$ \AA). Since these
values indicate a bound level very close to the asymptote (several MHz,
see fig.~\ref{scat}b),initially we were surprised that from our much less
accurate data (several 100 MHz) we can make such a good prediction of the
position of the last bound level. Obviously the large amount of observed data fixes the
accumulated phase of the wave function for this level along the
PEC up to $\approx 20$~\AA\ and the variation of the
binding energy with respect to the asymptote is restricted mainly
by the uncertainty of $C_6$. In addition, from figure~\ref{scat}b we see that
the smaller the binding energy, the less sensitive it is to the uncertainties
of the other long range parameters.

The inner wall of the potential ($R<3.5$ \AA), however, is not
well fixed by the experimental data. Could its change shift
the last level by several MHz without influencing the positions of
the other levels? The answer is negative. The wave functions for
small internuclear distances for the highly excited levels differ
only by a constant factor. For example, for a PEC with binding
energy $D_0-E_{40}\approx 3$~MHz for $v''=40$, the shapes of the wave
functions for $v''=40$ and $v''=35$ are almost identical up to 7.5
\AA\ and the ratio between them is:

\begin{equation}
  \frac{\Psi_{v''=35}(R)}{\Psi_{v''=40}(R)} \approx 35 \mbox{    for $R < 7.5$ \AA.}
\end{equation}

So, using the perturbation theory it is clear, that a change of
the potential for $R<3.5$ \AA, which will shift the position of
the level with $v''=40$ by 1 MHz will shift the level with
$v''=35$ by $\approx 1.2$ GHz
(0.04~\rcm). Since this change will not alter for example
the position of the lowest level ($v''=0$), the introduced shift
of the difference $E_{v''=35} - E_{v''=0}$ will exceed the experimental
uncertainty roughly by a factor of 6, which is a contradiction.

Due to the revised confidence interval for the $C_6$, the uncertainty of the
scattering length was not reduced compared to Ref.~\cite{Allard:02}
($112a_0 \div 800 a_0$), although the uncertainty of $C_6$ itself
is smaller than the assumed one in Ref.~\cite{Allard:02} ($\pm 5$ \% around the
theoretical prediction of $1.07\cdot 10^{7}$ \rcm\AA$^6$
\cite{Porsev:02}). Following the
dependence of the binding energy on the last level on $C_6$
(fig.~\ref{scat}b) we see that for smaller values of $C_6$ this
level approaches the asymptote, which makes an accurate
determination of the scattering length more difficult.

\section{Conclusions}
\label{concl}

The $^1$S+$^1$S molecular asymptote of the Ca dimer was studied
by employing the filtered laser excitation technique. The spectroscopic data
obtained in our previous study by Fourier transform spectroscopy were enriched
by adding 56 transition frequencies from ground state levels with $v''$ up to
38. This gave us an opportunity to reanalyze the existing description of the
long range part of the PEC from Ref.~\cite{Allard:02}.

By assuming the long range expansion (\ref{LR}) for the potential energy
curve to be valid already from 9.4 \AA,
the probability distributions of the dissociation
energy and the dispersion coefficients were derived from Monte Carlo analysis. Their
mean values and uncertainties are given in Table~\ref{param}. We showed, that
due to the large amount of experimental data, the long range analysis is almost
independent of the model functions used to describe the inner part of the PEC.

The reliability of the experimental potential for describing collisions
between two Ca atoms at low temperatures was checked by calculating the s-wave
scattering length and the position of the last bound level with respect to the
asymptote for the variety of long range extensions of the potential allowed by
the present experimental data.  Following the distribution of the long range
parameters, the binding energy of the last level was found to vary between
several 10 kHz and several MHz. This
confirmed our anticipations \cite{Allard:02} for a large and
positive scattering length, $250 a_0 < a < 1000 a_0$.
Contrary to the previous study, the present determination is purely experimental
and does not rely on the theoretical estimations for C$_6$.
Moreover, the difference in the values of C$_6$ from
Ref.~\cite{Porsev:02} and from this study exceeds the stated
uncertainties, although a small model dependence is still
existing by altering the connecting point between the inner
potential and the long range branch (compare fig.~\ref{10A}).

The Monte Carlo simulations open new perspectives for analyzing
the uncertainties of the parameters of the PEC. They could be
used also to determine the kind of spectroscopic data and their
required accuracy for achieving the desired uncertainty of the fitted
parameters. For example, we checked that including a vibrational
progression up to $v''=39$ with the present experimental accuracy
will not reduce the uncertainties of C$_6$ and D$_0$
significantly. This could be reached, however, even with the present
distribution of data, provided the accuracy can be improved.
Therefore, we believe that only high
resolution Doppler free spectroscopic techniques should be used
for further improvement on the Ca$_2$ ground state PEC,
especially at long internuclear distances.

\section{Acknowledgments}

This work is supported by DFG through SFB 407. The authors
appreciate the assistance of St. Falke during the experiments. A.
P. gratefully acknowledges the research stipend from the
Alexander von Humboldt Foundation.


 \end{document}